\documentclass[aps,prl,twocolumn,superscriptaddress]{revtex4-2}
\usepackage{graphicx}
\usepackage{dcolumn}
\usepackage{bm}
\usepackage{color}
\usepackage{hyperref}
\usepackage{amsmath}
\usepackage{verbatim}
\usepackage{docmute}

\begin{document}
\title{Magic-angle Twisted Bilayer Systems with Quadratic-Band-Touching:\\
Exactly Flat Bands with High-Chern Number} 
\author{Ming-Rui Li}
\affiliation{Institute for Advanced Study, Tsinghua University, Beijing 100084, China}
\author{Ai-Lei He}
 \email{heailei@yzu.edu.cn}
\affiliation{College of Physics Science and Technology, Yangzhou University, Yangzhou 225002, China}
\author{Hong Yao}
 \email{yaohong@tsinghua.edu.cn}
\affiliation{Institute for Advanced Study, Tsinghua University, Beijing 100084, China}
\affiliation{State Key Laboratory of Low Dimensional Quantum Physics, Tsinghua University, Beijing 100084, China}
\date{\today}
\begin{abstract}
Studies of twisted moiré systems have been mainly focused on two-dimensional (2D) materials such as graphene with Dirac points and transition-metal-dichalcogenide so far. Here we propose a twisted bilayer of 2D systems which feature stable quadratic-band-touching points and find exotic physics different from previously studied twisted moiré systems. Specifically, we show that exactly flat bands can emerge at magic angles and, more interestingly, \textit{each} flat band exhibits a high Chern number ($C=\pm 2$). We further consider the effect of Coulomb interactions in such magic-angle twisted systems and find that the ground state supports the quantum anomalous Hall effect with quantized Hall conductivity $2\frac{e^2}{hc}$ at certain filling. Furthermore, the possible physical realization of such twisted bilayer systems will be briefly discussed. 
\end{abstract}

\maketitle

{\bf Introduction:} 
Twisted moiré systems, especially twisted bilayer graphene (TBG), have attracted enormous attention in recent years due to the emergence of topological flat bands and various interesting phases such as correlation insulators and unconventional superconductivity \cite{Cao20181,Cao20182,Kerelsky2019,Xie2019,Sharpe_2019,Jiang2019,doi:10.1126/science.aay5533,Choi2019,Lu_2019,Yankowitz_2019,PhysRevLett.124.076801,choi2020tracing,wong2020,Zondiner2020,Saito2020,Stepanov2020,Arora2020,Nuckolls2020,Saito_2021,Liu_2021,Wu_2021,Das_2021,Park_2021,Saito_2021,Rozen_2021,lu2020multiple,xie2021fractional}. Since its experimental discovery, extensive studies of such systems have been done on both experimental and theoretical sides. The theoretical prediction of flat bands in TBG was 
made by Bistritzer and MacDonald (BM) \cite{Bistritzer12233}, in their paper the BM Hamiltonian and the moiré band theory were developed to study TBG and other twisted moiré systems. Furthermore, a generalization of the BM model was developed \cite{PhysRevB.103.205411} and a more complete description and understanding of the flat bands in a twisted bilayer system were obtained through perturbation theory. Based on the moiré band theory, enormous numbers of studies were done to explore the topological features \cite{PhysRevLett.123.036401, PhysRevB.99.195455, PhysRevX.9.021013, PhysRevB.103.205412} as well as the interaction effects \cite{PhysRevLett.121.087001,PhysRevLett.121.217001,PhysRevLett.121.257001,PhysRevX.8.031087,PhysRevX.8.031089,PhysRevX.8.041041,PhysRevB.97.235453,PhysRevB.98.075109,PhysRevB.98.075154,PhysRevB.98.081102,PhysRevB.98.121406,Guinea13174,PhysRevB.98.241407,PhysRevB.98.245103,You2019,PhysRevB.99.121407,PhysRevLett.122.026801,PhysRevLett.122.246401,PhysRevLett.122.246402,PhysRevLett.122.257002,PhysRevLett.123.157601,PhysRevLett.123.197702,PhysRevLett.123.237002,HUANG2019310,PhysRevLett.124.046403,PhysRevLett.124.097601,PhysRevLett.124.166601, PhysRevLett.124.167002,PhysRevLett.124.187601,PhysRevResearch.2.023237,PhysRevResearch.2.023238,PhysRevB.101.060505,PhysRevB.102.205111,PhysRevB.102.035136,PhysRevLett.125.257602,PhysRevX.10.031034,Christos29543,PhysRevX.11.011014,PhysRevB.103.035427, PhysRevB.103.205413, PhysRevB.103.205414, PhysRevB.103.205415, PhysRevB.103.205416,PhysRevResearch.3.013033,PhysRevResearch.3.013242,Khalaf_2021,PhysRevB.103.235401,wang2021exact} of TBG systems; non-trivial topology of the flat bands has been shown, and huge progress has been made in understanding the interacting phases. 

Although twisted systems have attracted vast research attention, studies of them have been mainly limited to twisted graphene systems with Dirac fermions and twisted transition-metal-dichalcogenide (TMD) \cite{PhysRevLett.121.026402,PhysRevLett.122.086402,Tang2020,Regan2020,PhysRevB.102.201104,Shabani2021,PhysRevB.103.155142,Xu2020,Bi2021,PhysRevB.103.L241110,PhysRevB.104.075150,Zhang2020,li2021quantum,devakul2021magic,devakul2021quantum,kiese2021tmds};  explorations of twisted systems with other types of fermions, such as those with quadratic band touchings, remain scarce. It is desired to study such new types of twisted systems mainly for the following reasons. On the one hand, the larger density of states in these systems may lead to nontrivial interacting phases \cite{SYFK,Vafek-KYang-PRB-2010,MacDonald-PRB-2010,YZYou-Fradkin-PRB-2013}. On the other hand, the possibility of realizing higher-Chern number flat bands in such twisted systems is attractive as high-Chern-number flat bands can provide an arena to realize various exotic fractional quantum Hall effects \cite{PhysRevLett.106.236804, PhysRevLett.107.126803, PhysRevB.84.241103,PhysRevB.86.201101,PhysRevB.86.241112,PhysRevLett.110.106802,PhysRevB.101.235312,PhysRevLett.126.026801,PhysRevB.104.125107} and its realization in quantum materials remains elusive. 

\begin{figure}[b]
\centering
  \includegraphics[width=0.95\linewidth]{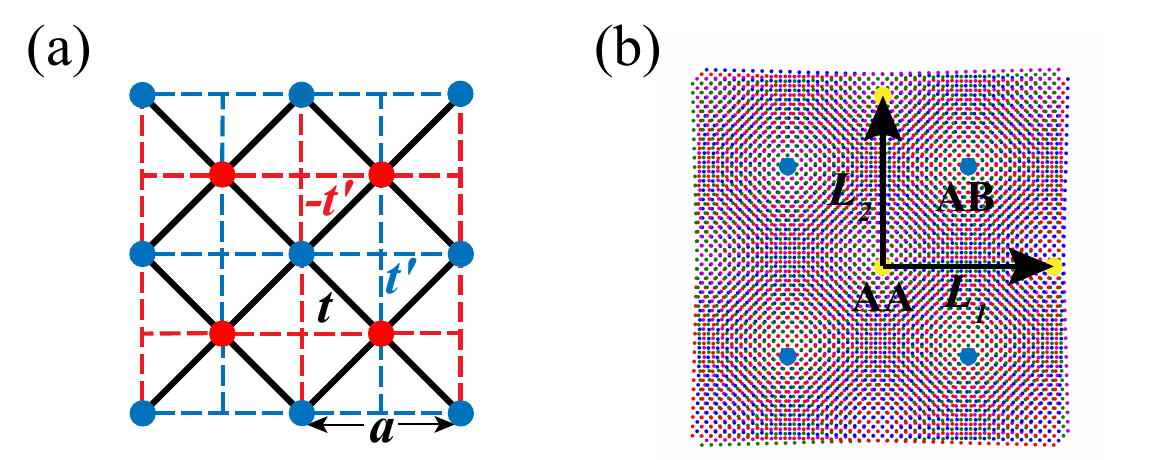}
  \caption{(a) Schematic representation of the checkerboard lattice. Blue and red sites constitute sublattices $A$ and $B$, respectively. Note that the intra-sublattice hopping in sublattice $A$ is opposite to that in sublattice $B$. (b) The moiré pattern of the TBCB lattice.}
  \label{fig:cklattice}
\end{figure}

In this paper, we investigate a twisted bilayer of systems with quadratic band touching, with focus on the twisted bilayer checkerboard (TBCB) model. The checkerboard lattice model in each single layer was proposed by Sun, Yao, Fradkin, and Kivelson (SYFK) \cite{SYFK} to realize a stable quadratic band touching point (QBTP) Note that for $AB$-bilayer graphene, the putative QBTP is not stable in the presence of trigonal hopping \cite{RevModPhys.81.109}. We found that such twisted systems can host two exactly flat bands per spin in the chiral limit, and more interestingly, \textit{each} flat band has nontrivial topology with high Chern number $C=\pm2$. 
Note that, in contrast to TBG with a total of eight flat bands, there 
are only four flat bands in the TBCB. 
In the presence of Coulomb interactions, by projecting them onto the topological flat bands of $C=\pm 2$ in TBCB systems similar to the analysis employed for TBG \cite{PhysRevB.103.205413, PhysRevB.103.205414}, we showed that the interaction prefers the ground state with minimum Chern number; at charge-neutrality ($\nu=0$) the ground state is an insulator with Chern number $C=0$, while for $\nu=\pm 1$ the ground state possesses Chern number $\pm2$ and exhibits the quantum anomalous Hall effect \cite{Sharpe_2019,Stepanov2020,doi:10.1126/science.aay5533,doi:10.1126/science.1234414}. We further propose a possible optical-lattice realization of the TBCB with topological flat bands, providing a promising route to study various correlated phases in TBCB experimentally.

{\bf Quadratic-band-touching model:} 
One prototype model hosting stable quadratic-band-touching points (QBTPs) is the checkerboard model proposed by SYFK \cite{SYFK}. As shown in Fig. \ref{fig:cklattice}(a), it can be described by the tight-binding Hamiltonian:
    $H = -\sum_{i,j}t_{ij}c^{\dagger}_{i}c_{j}$,
with the hopping amplitude $t_{ij}$ between sites $i$ and $j$. Here we consider nearest-neighbor hopping $t$ and next-nearest-neighbor hopping $\pm t'$, as shown in Fig.~\ref{fig:cklattice}(a). Note that the lattice consists of two sublattices labeled as $A$ and $B$. By performing the Fourier transformation $c_{i\in A(B)}=\frac{1}{\sqrt{N}}\sum_k e^{i\mathbf{k}\cdot \mathbf{r_i}}\psi_{\mathbf{k},A(B)}$, where $N$ is the number of unit cells, we obtain $H_0=\sum_{\mathbf{k}} \psi^\dag(\mathbf{k}) H_0(\mathbf{k})\psi(\mathbf{k})$ with $\psi^\dag=(\psi^\dag_A,\psi^\dag_B)$. $H_0(\mathbf{k})$ is the two-band Bloch Hamiltonian:
$H_0\left(\mathbf{k}\right) = d_x(\mathbf{k})\sigma_x + d_z(\mathbf{k})\sigma_z$,
where $d_z(\mathbf{k})= -4t \cos \left(k_x/2\right) \cos \left(k_y/2\right)$ and $d_x(\mathbf{k})=2t'(\cos k_{x} - \cos k_{y})$ \footnote{Notice that we've adopted the real gauge such that the $d_i(\mathbf{k})$ are real, in this case, the Hamiltonian is not periodic and satisfies $H(\mathbf{k} + \mathbf{G}_i) = \sigma_z H(\mathbf{k})\sigma_z$, where $\mathbf{G}_i$ is the reciprocal vector of the lattice\cite{PhysRevX.9.021013}. We can also choose a periodic and complex gauge and in this case, we have $d_{x}(\mathbf{k})-i d_{y}(\mathbf{k})=-\left(1+e^{-i k_{x}}+e^{-i k_{y}}+e^{-i\left(k_{x}+k_{y}\right)}\right)$.}. It is straightforward to obtain the dispersion of two bands: $\epsilon_{\mathbf{k},\pm}=\pm\sqrt{d^2_x(\mathbf{k})+d^2_z(\mathbf{k})}$. By expanding the periodic Bloch Hamiltonian around $\mathbf{M} = (\pi,\pi)$ where two bands cross (namely $\mathbf{k}\to \mathbf{k}+\mathbf{M}$) and keeping only the lowest orders in $\mathbf{k}$, we obtain
\begin{equation}
    H_0\left(\mathbf{k}\right)=t k_x k_y \sigma_x+t'(k_x^2-k_y^2)\sigma_z.
    \label{equ:cbHamiltonian}
\end{equation}
The dispersion around $\mathbf{M}$ is quadratic and $\mathbf{M}$ is a called the quadratic-band-touching point (QBTP). To transform the Hamiltonian into a form with explicit chiral symmetry, we can perform a basis transformation $\psi\to U\psi$ with $U=e^{i\frac{\pi}{2}\sigma^x}$ and obtain $\tilde H_0(\mathbf{k})=U^\dag H_0(\mathbf{k})U$ as 
\begin{equation}
\tilde H_0\left(\mathbf{k}\right)=t k_x k_y \sigma_x + t'(k_x^2-k_y^2)\sigma_y. \label{equ:chiral}
\end{equation}
Note that the QBTP features a Berry phase of $2\pi$, which is twice of that of a Dirac point. Hereafter, unless stated otherwise, we shall assume $t'=t/2$, so that the dispersion $E_\pm(\mathbf{k})=\pm \frac{t}{2} \mathbf{k}^2$. 

\begin{figure}[t]
\centering
  \includegraphics[width=0.95\linewidth]{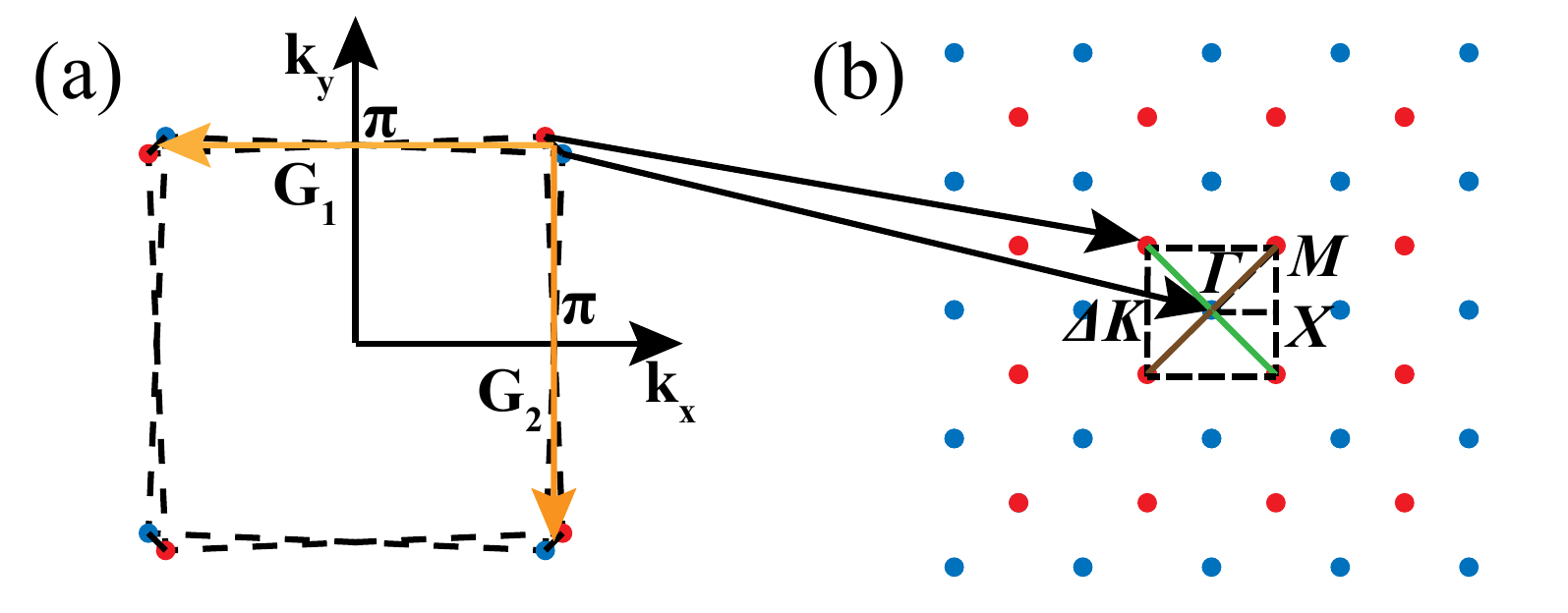}
  \caption{
  (a) The Brillouin zone of the top (blue) and bottom (red) layer. $\mathbf{G}_1$ and $\mathbf{G}_2$ is plotted in yellow. (b) The moiré Brillouin zone of the TBCB lattice.  The blue points represent the $\mathbf{M}$ points of the top layer while the red points represent the $\mathbf{M}$ points of the bottom layer. We pick the square surrounded by dashed lines as our mBZ, and label the three high-symmetry points as $\Gamma$, $X$, and $M$.}
  \label{fig:morielattice}
\end{figure}

{\bf Exactly flat bands at magic angles:} 
It is desirable to investigate novel physics in a twisted bilayer of systems with quadratic-band-touching fermions for the exotic property of the QBTP\cite{SYFK}. In this paper, we consider the twisted bilayer of the checkerboard lattice, and explore its physics such as totally flat bands at magic angles and the high Chern number of those flat bands. The lattice structure of the TBCB lattice is shown in Fig.~\ref{fig:cklattice}(b). 

Here we mainly focus on the low-energy physics of the twisted bilayer system with quadratic band touching by employing the continuum model describing the low-energy band structure around the QBTP $\mathbf{M}$. Using the moiré band theory introduced by Bistritzer and MacDonald \cite{Bistritzer12233}, we obtained the inter-layer hopping matrices:  $T^{\alpha\beta}_{\mathbf{p}\mathbf{p}^{\prime}} =\sum_{\mathbf{G}_{1} \mathbf{G}_{2}} \frac{t_{\mathbf{p}+\mathbf{G}_{1}}}{\Omega} 
e^{i \mathbf{G}_{1} \cdot \mathbf{\tau}_{\alpha}-i \mathbf{G}_{2}\cdot \mathbf{\tau}_{\beta} } \delta_{\mathbf{p}+\mathbf{G}_{1}, \mathbf{p}^{\prime}+ M_{\theta}\mathbf{G}_{2}}, $
where $\mathbf{G}_{1}, \mathbf{G}_{2}$ are the reciprocal vectors of the lattice, $\alpha$ and $\beta$ labels the sublattice indices $A$ and $B$, respectively, $M_\theta$ represents the rotation by angle $\theta$, and $\mathbf{\tau}_{\alpha (\beta)}$ represents the relative coordinates of the sublattice $\alpha$$(\beta)$ in the unit cell. For the checkerboard lattice shown in Fig. 1(a), we have $\mathbf{\tau}_{A} = (0,0)$ and $\mathbf{\tau}_{B} = (\frac{1}{2},\frac{1}{2})$ in the unit of lattice constant $a$. Inspired by the TBG theory, we only keep the largest four $t_{\mathbf{p}}$ terms, i.e, the terms with $\mathbf{p} - \mathbf{p}^{\prime}=C_{4z}^{i}(M_{\theta}-1)\mathbf{M}$, where $i=0,1,2,3$ . With these four hoppings we can construct the moiré Brillouin zone (mBZ) as shown in Fig.~\ref{fig:morielattice}(b) and the hopping matrices take the form:
\begin{equation}
T_{1}=\left(
\begin{array}{cc}
w_{AA} & w_{AB} \\
w_{AB} & w_{AA} \\
\end{array}
\right),~~~ 
T_{2}=\left(
\begin{array}{cc}
w_{AA} & -w_{AB} \\
 -w_{AB} & w_{AA} \\
\end{array}
\right),
\end{equation}
where $T_{1}(T_2)$ is the hopping matrix of hopping along the green (brown) lines in the mBZ as shown in Fig.~\ref{fig:morielattice}(b) \footnote{Note that $T_{1,2}=w_AA\sigma_0\pm w_{AB}\sigma_x$, which commute with the unitary transformation $U=e^{i\frac{\pi}{2}\sigma_x}$, such that the tunneling matrix is not changed by the unitary transformation}. 

\begin{figure}[t]
\centering
  \includegraphics[width=1.0\linewidth]{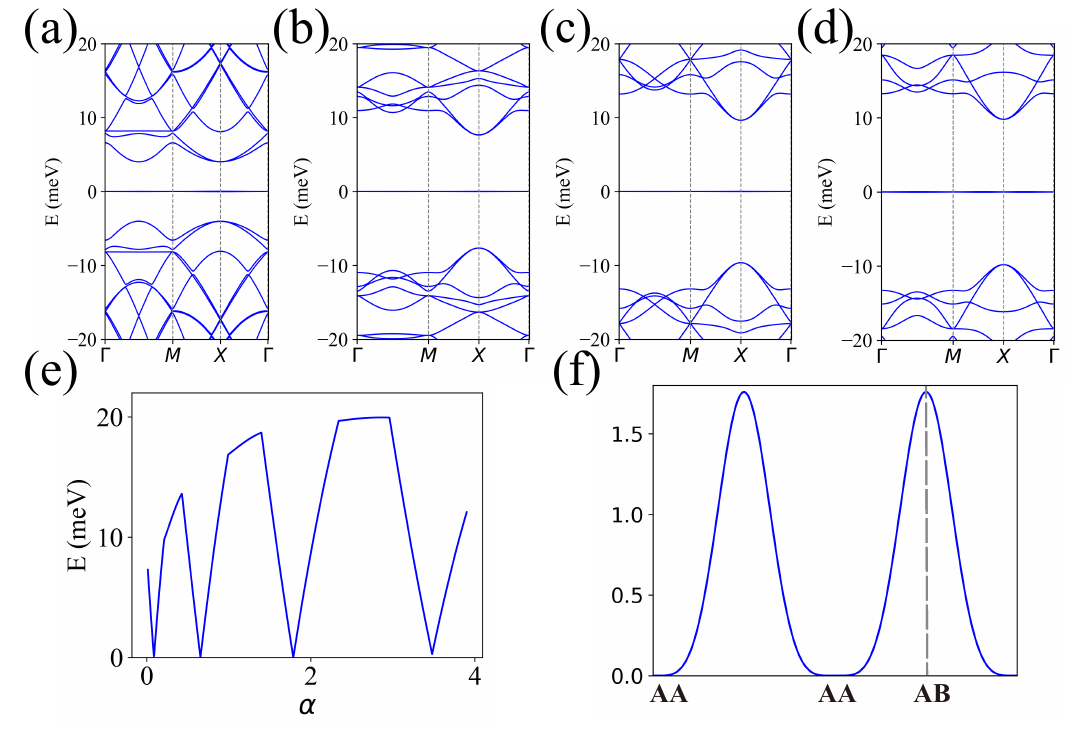}
  \caption{(a)-(d) The band structure of the TBCB lattice with $\alpha=0.26$, $2.16$, $5.93$, and $11.62$, respectively. For each case, there exist two degenerate and totally flat bands in the middle. (e) The bandwidth of the middle two bands. The bandwidth drops to exactly zero at the magic angles. (f) The wavefunction density for a single zero mode at QBTP $\mathbf{M}$ along the line $AA-AB$, a zero point exists at the $AA$ stacking point.}
  \label{fig:ckflatbands}
\end{figure}

Assuming the chiral limits $w_{AA}=0$ and $w_{AB}=w$\cite{PhysRevLett.122.106405}, we numerically computed the moiré bands and observed exactly flat bands 
for a series of magic angles as shown in Fig.~\ref{fig:ckflatbands}. The band structure is controlled by a single parameter $\alpha=\frac{w}{t' k_{\theta}^2}=\frac{wa^2}{8t'\pi^2 \sin^2(\theta/2)}$ which is proportional to $1/\sin^2(\theta/2)$. Note that the parameter $\alpha$ is qualitatively different from its counterpart $\frac{w}{\sin(\theta/2)}$ for TBG \cite{PhysRevLett.122.106405}. As a consequence, the magic angle for the TBCB can be much larger compared with TBG. This property makes the twist angle of the TBCB system easier to be tuned experimentally.

{\bf Origin of the exactly flat bands:}
We now provide an analytical understanding of the origin of exactly flat bands at those magic angles. First, we perform the Fourier transformation and obtain the hopping matrices in real space
$T(\mathbf{r})=\sum_{n=1}^{2} 2T_{n} \cos( \mathbf{q}_{n}\cdot \mathbf{r})$,
where $\mathbf{q}_{1}=\frac{k_{\theta}}{\sqrt{2}}(1, 1)$ and $\mathbf{q}_{2}=\frac{k_{\theta}}{\sqrt{2}}(1, -1)$ with  $k_{\theta}=\frac{2\sqrt{2}\pi}{a}\sin\frac{\theta}{2}$. Since the system preserves the chiral symmetry when $w_{AA}=0$, we choose the basis $\Phi(\mathbf{r})=\left(\psi_{1,A}, \psi_{2,A}, \psi_{1,B}, \psi_{2,B}\right)^{\mathrm{T}}$, where $1$ and $2$ are the layer indices and $A$ and $B$ are the sublattice indices, such that the Hamiltonian is given by
\begin{equation}
H(\mathbf{r})=\left(
\begin{array}{cc}
0 & \mathcal{D}(\mathbf{r}) \\
\mathcal{D}^*(-\mathbf{r}) &0  \\
\end{array}
\right), 
\end{equation}
where
$D(\mathbf{r})$ is completely antiholomorphic:
\begin{eqnarray}
&&\mathcal{D}(\mathbf{r})=\nonumber\\
&&\left(
\begin{array}{cc}
-i\frac{t}{2} \overline{\partial}^2 & 2w[\cos(\mathbf{q}_1\!\cdot\!\mathbf{r}) \!-\! \cos(\mathbf{q}_2\!\cdot\!\mathbf{r})] \\
 2w[\cos(\mathbf{q}_1\!\cdot\!\mathbf{r}) \!-\! \cos(\mathbf{q}_2\!\cdot\!\mathbf{r})] &-i\frac{t}{2} \overline{\partial}^2  \\
\end{array}
\right),~~~ 
\end{eqnarray}
with $\overline{\partial}\equiv \partial_{\bar z}=\partial_x-i\partial_y$ [hereinafter we shall use $\mathbf{r}=(x,y)$ and $z=x+iy$ interchangeably]. 

The QBTP at $\mathbf{M}$ with zero energy is protected by symmetry. Explicitly, the zero-energy wave function  $\phi_\mathbf{M}$ satisfying $H\phi_\mathbf{M}=0$ is given by $\phi_\mathbf{M}^T=(0,0,\psi^T_\mathbf{M})$, where $\psi_\mathbf{M}(\mathbf{r})$ is a two-component wavefunction satisfying $\mathcal{D}\psi_\mathbf{M}(\mathbf{r})=0$. 
Since $\mathcal{D}$ is antiholomorphic, we can construct
the wavefunction $\psi_{\mathbf{k}}(\mathbf{r})=f_\mathbf{k}(z)\psi_\mathbf{M}(\mathbf{r})$, where $f_\mathbf{k}(z)$ is a holomorphic function, with the following feature: $\mathcal{D}\psi_\mathbf{k}(r)=f_\mathbf{k}(z)\mathcal{D}\psi_\mathbf{M}(\mathbf{r})=0$. If such a holomorphic function $f_\mathbf{k}(z)$ exists for every $\mathbf{k}$ in the mBZ, the wavefunction of the totally flat band with momentum $\mathbf{k}$ is obtained.
Note that $\psi_{\mathbf{M}}$ satisfy the Moire boundary condition $\psi_{\mathbf{M}}\left(z+L_j\right)=\sigma_z\psi_{\mathbf{M}}(z)$ which means $f_{\mathbf{k}}\left(z+L_j\right)=e^{i \mathbf{k}\cdot \mathbf{L}_{j}} f_{\mathbf{k}}(z)$, where $L_{j}=\mathbf{L}_{j}\!\cdot\!\hat x+i \mathbf{L}_{j}\!\cdot\! \hat y$ \cite{PhysRevLett.122.106405}; 
consequently, $f_{\mathbf{k}}(z)$ must have a simple pole, and such a construction fails in general. However, when $\psi_{\mathbf{M}}(\mathbf{r})$ has a zero point at special twist angles, nonsingular $f_{\mathbf{k}}(z)$ is permitted, and the exactly flat bands from the construction above can exist. Such a special twist angle at which $\psi_{\mathbf{M}}(\mathbf{r})$ has zeros is a so-called magic angle. Indeed, our calculation of $\psi_\mathbf{M}$ with zero energy at the magic angle shows that $\psi_\mathbf{M}(\mathbf{r})$ is zero when $\mathbf{r}$ is at AA stacking point, as shown in Fig.~\ref{fig:ckflatbands}(f).

{\bf Derivation of the first magic angle:}  Based on the requirement that at the magic angle the wave function $\phi_\mathbf{M}(\mathbf{r})=0$ for $\mathbf{r}$ at the $AA$ stacking point, we can analytically derive the parameter $\alpha$ corresponding to the magic angle. Solving the equation $\mathcal{D}\psi_{\mathbf{M}}(\mathbf{r})=0$ perturbatively in the parameter $\alpha<1$, we obtain the spinor wave function
\begin{equation}
\psi_\mathbf{M}(\mathbf{r})=(1 + u_{1} \alpha +u_{2} \alpha^{2} +\cdots)\frac{1}{\sqrt{2}}
\left(\begin{array}{c}
 1\\
\pm 1
\end{array}\right),
\end{equation}
where $\cdots$ represents higher-order terms in $\alpha$ and $u_j(\mathbf{r})$ carries momentum $m\mathbf{q}_{1} + n\mathbf{q}_{2}$ with $|m|+|n|=j$. Up to the second order $u_2$, one can get the solution $u_1(\mathbf{r})=\mp2\alpha\left[\cos(\mathbf{q}_1\!\cdot\!\mathbf{r}) \!+\! \cos(\mathbf{q}_2\!\cdot\!\mathbf{r})\right]$ and $u_2(\mathbf{r})=\frac{1}{2}\alpha^2\left[\cos(2\mathbf{q}_1\!\cdot\!\mathbf{r}) \!+\! \cos(2\mathbf{q}_2\!\cdot\!\mathbf{r})\right]$. Requiring that the wavefunction is zero at the $AA$ stacking point, namely $\psi_\mathbf{M}(\mathbf{0})=(0,0)^T$, we obtain the first magic angle solution $\alpha=\alpha_0=2-\sqrt{3}\approx 0.268$, which is very close to the numerically-obtained first magic angle shown in Fig. \ref{fig:ckflatbands}(a).

Another (probably more intuitive) way to derive the magic angle is to require the vanishing of 
the inverse effective mass of the fermions at the QBTP $\mathbf{M}$ which is qualitatively different from the TBG system, which only requires the vanishing of the Fermi velocity at magic angles. Requiring the vanishing of the inverse effective mass of the fermions at the QBTP $\mathbf{M}$, we obtain the first magic-angle parameter $\alpha=\frac{1}{\sqrt{12}}\approx0.289$, which is also quite close to the value obtained numerically in Fig.~\ref{fig:ckflatbands}(a). Details of computing the inverse effective mass are shown in Appendix B. 

\begin{figure}[t]
\centering
  \includegraphics[width=4.5cm]{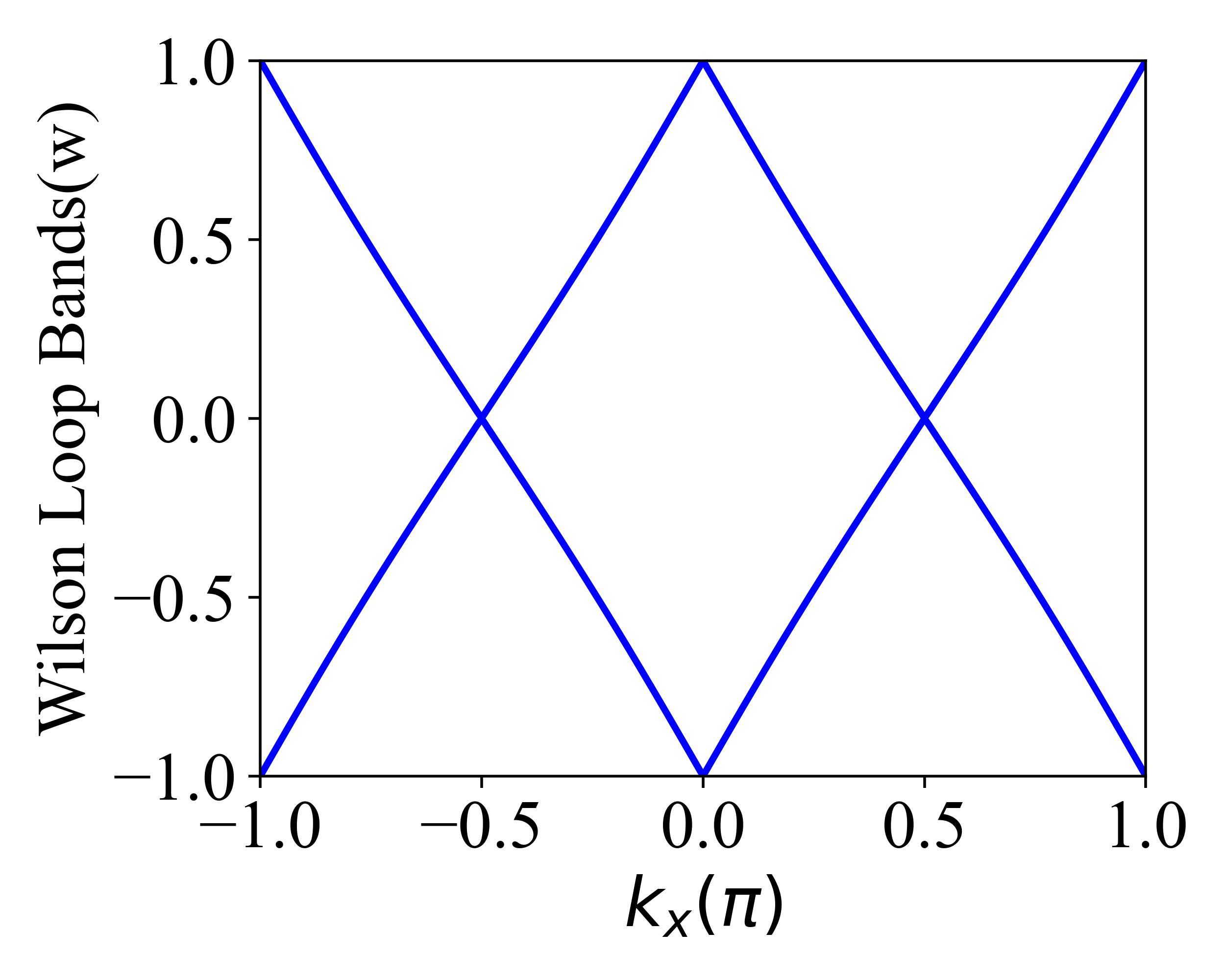}
  \caption{Wilson-loop bands of the flat bands at the first magic angle $\alpha=0.26$. The two Wilson loops wind from $\pm1$ to $\mp1$ twice as $k_1$ varies from $-\pi$ to $\pi$. Thus the winding number of the Wilson loop is $\pm2$.}
  \label{fig:wilsonloop}
\end{figure}

{\bf High-Chern number of exactly flat bands:} 
To analyze the topology of these flat bands, we first calculate the Wilson loop's winding number of the TBCB lattice. 
For the Bloch states in the moiré Brillouin zone, to define the Wilson loop, we need to first restore the periodicity of the Bloch states and thus need to introduce the extra embedding matrix $V_{\mathbf{Q}, \mathbf{Q}^{\prime}}^{\mathbf{G}}=\delta_{\mathbf{Q}-\mathbf{G}, \mathbf{Q}^{\prime}}$. Consequently, the Wilson loop for the TBCB lattice is calculated by
\begin{equation}
W\left(k_{1}\right)=U_{k_{1}, 0}^{\dagger} U_{k_{1}, \frac{2 \pi}{N}}  \cdots U_{k_{1}, \frac{(N-1) \times 2 \pi}{N}}^{\dagger} V^{\mathbf{G}} U_{k_{1}, 0}. 
\end{equation}
where $U_{k_1,k_2}=U_{\mathbf{k}}=\left(\left|u_{1 \mathbf{k}}\right\rangle, ...,\left|u_{N} \mathbf{k}\right\rangle\right)$ with $\mathbf{k}=(k_1,k_2)$ \cite{PhysRevX.6.021008,PhysRevLett.123.036401,PhysRevB.103.205412}. We assume that $\mathbf{b_1}$ and $\mathbf{b_2}$ are the reciprocal vectors of the moiré lattice, $\mathbf{k} = k_1 \mathbf{b_1} + k_2 \mathbf{b_2}$. We keep $k_1$ unchanged and vary $k_2$ to obtain the flow of the Wilson loop spectrum along $k_1$. We find that the Wilson loop winds from $\pm1$ to $\mp1$ twice while $k_1$ goes from $-\pi$ to $\pi$, so the winding of the Wilson loop at the first magic angle is $\pm2$, as shown in the Fig. \ref{fig:wilsonloop}. This is also true for other magic angles that we identified. When adding more bands into the Wilson loop calculation such as the middle six bands, the winding is still preserved which suggests that the topology is stable. In fact, the anti-unitary particle-hole symmetry $P$ (see the Appendix C) protects the degeneracy of the Wilson bands at $k_1=0/\pi$ in the same way as in TBG \cite{PhysRevB.103.205412}.

In the non-interacting limit, the degeneracy of the two flat bands for the TBCB is protected by the time-reversal symmetry $T$. To give a simple illustration of the topology with the existence of interactions, we apply a weak0time-reversal-symmetry-breaking term which preserves all other symmetries  except the mirror symmetry:  $A\sigma_{y}\sin(\frac{k_x}{2})\sin(\frac{k_y}{2})$. The degeneracy is then lifted, while the flatness of bands is still well preserved. Assuming $A/w_{AB}=0.5$, we have calculated the Berry curvature of the two flat bands and the corresponding Chern number: $C_n = \frac{1}{2\pi}\int_{\text{mBZ}} \mathcal{B}_{n}d_{k_x}d_{k_y}.$
Our result shows that the lower band hosts a Chern number of $\mp2$, while the upper band host a Chern number of $\pm2$ where the sign of the Chern number depends on the sign of $A$. For a spinless TBCB system, this is the only possible way for the two bands to split; the topology of the system is highly nontrivial with high Chern number $C=\pm2$. It is noteworthy that this high Chern number of $\pm 2$ is realized in the flat bands of a twisted bilayer systems \footnote{High-Chern number flat bands can also be realized in twisted graphene multilayers \cite{ledwith2021family,wang2021hierarchy}} with stable QBTPs, and in our system, the QBTPs have only one valley, which will lead to different physics.

{\bf Correlation effect:} Interactions can play an essential role in the physics of twisted bilayer systems\cite{PhysRevLett.121.087001,PhysRevLett.121.217001,PhysRevLett.121.257001,PhysRevX.8.031087,PhysRevX.8.031089,PhysRevX.8.041041,PhysRevB.97.235453,PhysRevB.98.075109,PhysRevB.98.081102,PhysRevB.98.121406,Guinea13174,PhysRevB.98.241407,PhysRevB.98.245103,You2019,PhysRevLett.122.026801,PhysRevLett.122.246401,PhysRevLett.122.246402,PhysRevLett.122.257002,PhysRevLett.123.157601,PhysRevLett.123.197702,PhysRevLett.123.237002,HUANG2019310,PhysRevLett.124.046403,PhysRevLett.124.097601,PhysRevLett.124.166601, PhysRevLett.124.167002,PhysRevLett.124.187601,PhysRevResearch.2.023237,PhysRevResearch.2.023238,PhysRevB.101.060505,PhysRevB.102.205111,PhysRevB.102.035136,PhysRevLett.125.257602,PhysRevX.10.031034,Christos29543,PhysRevX.11.011014,PhysRevB.103.035427, PhysRevB.103.205413, PhysRevB.103.205414, PhysRevB.103.205415, PhysRevB.103.205416,PhysRevResearch.3.013033,PhysRevResearch.3.013242,Khalaf_2021,PhysRevB.103.235401}. Here, 
we consider 
the Coulomb interactions: 
\begin{equation}
\mathcal{H}_{I}=\frac{1}{2 A} \sum_{\mathbf{G} \in \mathcal{G},\mathbf{q} \in \mathrm{mBZ}} V(\mathbf{q}+\mathbf{G}) \delta \rho_{-\mathbf{q}-\mathbf{G}} \delta \rho_{\mathbf{q}+\mathbf{G}},
\end{equation}
where $\delta\rho_\mathbf{q}=\sum_\mathbf{r} e^{i\mathbf{q}\cdot \mathbf{r}}(\rho_\mathbf{r}-\frac{1}{2}\delta_{\mathbf{q,0}}\delta_{\mathbf{G,0}})$, $\mathcal{G}$ represents the moiré reciprocal lattice vectors in the Brillouin zone (BZ) of the original lattice, and $V(\mathbf{q})=\pi d^{2} U_{d} \frac{\tanh (d|\mathbf{q}|/2)}{d|\mathbf{q}|/2}$ is the screened Coulomb potential with $U_{d}=e^2/(\epsilon d)$ \cite{PhysRevB.103.205413}; $d$ is the distance between the TBG and the top or bottom gate, and $\rho_\mathbf{r}$ is the charge density at $\mathbf{r}$. To solve its low-energy physics, one can project the Hamiltonian onto the subspace of the two flat bands. To do so, we employ fermion operators in the mBZ energy band basis $c_{n,s}^{\dagger}(\mathbf{k})=\sum_{\mathbf{Q} \alpha} u_{\mathbf{Q}, \alpha, n}(\mathbf{k}) f_{\alpha,s}^{\dagger}(\mathbf{k} + \mathbf{Q})$, where $\mathbf{Q}\in\mathcal{Q}_{\pm}$ and $\mathcal{Q}_{\pm}$ is the collection of the sites of layer $l=\pm$ of the mBZ as plotted in Fig.~\ref{fig:morielattice}. Here, $n$ is the moiré band index, and $n=\pm1$ represents the two flat bands. Due to its nontrivial topology, we cannot define a symmetric smooth and periodic wave function $u_{\mathbf{Q}, \alpha, n}(\mathbf{k})$ \cite{PhysRevX.10.031034}. Here we adopt a periodic gauge that satisfies $u_{\mathbf{Q}, \alpha, n}\left(\mathbf{k}+\mathbf{b}_{i}\right)=u_{\mathbf{Q}-\mathbf{G}, \alpha, n}(\mathbf{k})$. 
Similar to the treatment of interactions in TBG \cite{PhysRevB.103.205413,PhysRevB.103.205414}, after the projection into the flat band subspace the interacting Hamiltonian 
for the TBCB system is written as 
\begin{eqnarray}
H_{I}=\frac{1}{2 \Omega_{\text {tot }}} \sum_{\mathbf{q} \in \text { mBZ }} \sum_{\mathbf{G} \in \mathcal{G}} O_{-\mathbf{q},-\mathbf{G}} O_{\mathbf{q}, \mathbf{G}},
\end{eqnarray}
where $\Omega_{\text {tot }}$ is the total area of the TBCB system,
\begin{eqnarray}
O_{\mathbf{q}, \mathbf{G}}=\sum_{\mathbf{k} s} \sum_{m, n=\pm 1} \sqrt{V(\mathbf{q}+\mathbf{G})} M_{m, n}(\mathbf{k}, \mathbf{q}+\mathbf{G}) \nonumber\\
~~~~~~\times\left(c_{m, s}^{\dagger}(\mathbf{k}+\mathbf{q}) c_{n, s}(\mathbf{k})-\frac{1}{2} \delta_{\mathbf{q}, 0} \delta_{m, n}\right),
\end{eqnarray}
and
\begin{eqnarray}
M_{mn}(\mathbf{k}, \mathbf{q}\!+\!\mathbf{G})\!=\!\sum_{\alpha,\mathbf{Q} \in \mathcal{Q}_{\pm}} u_{\mathbf{Q}-\mathbf{G}, \alpha, m}^{*}(\mathbf{k}+\mathbf{q}) u_{\mathbf{Q}, \alpha, n}(\mathbf{k}).~~~~~
\end{eqnarray}
As discussed in the Appendix C, considering 
the $C_{2z}$, $T$, and $P$ symmetries, the $M$ matrix is constrained as follows:
\begin{equation}
M(\mathbf{k}, \mathbf{q}+\mathbf{G})=\zeta^{0} \alpha_{0}(\mathbf{k}, \mathbf{q}+\mathbf{G}) + i \zeta^{y} \alpha_{2}(\mathbf{k}, \mathbf{q}+\mathbf{G}),\label{equ:Mmatrix}
\end{equation}
where $\zeta$ represents the Pauli matrix for the two-flat-band subspace, $\alpha_0(\mathbf{k}, \mathbf{q}+\mathbf{G})$ and $\alpha_{2}(\mathbf{k}, \mathbf{q}+\mathbf{G})$ are real numbers with the constraints $\alpha_{0}(\mathbf{k}, \mathbf{q}+\mathbf{G})=\alpha_{0}(\mathbf{k}+\mathbf{q},-\mathbf{q}-\mathbf{G})$ and $\alpha_{2}(\mathbf{k}, \mathbf{q}+\mathbf{G})=-\alpha_{2}(\mathbf{k}+\mathbf{q},-\mathbf{q}-\mathbf{G})$ and $\alpha_{a}(\mathbf{k}, \mathbf{q}+\mathbf{G})=\alpha_{a}(-\mathbf{k},-\mathbf{q}-\mathbf{G})$ for $a=0,2$ (see details in the Appendix C). 

In the Chern band basis \cite{PhysRevB.103.205412,PhysRevB.103.205413} which is the eigenstate of the flat bands with Chern number $C=2e$:
$d_{\mathbf{k}, e, s}^{ \dagger}=\frac{c_{1, s}^{\dagger}(\mathbf{k})+i e c_{-1, s}^{\dagger}(\mathbf{k})}{\sqrt{2}}$, 
where $e=\pm1$, we can rewrite the operator $O_{\mathbf{q}, \mathbf{G}}$ in a diagonal way
\begin{equation}
\begin{aligned}
O_{\mathbf{q}, \mathbf{G}}&=\sum_{\mathbf{k} s} \sum_{e=\pm 1} \sqrt{V(\mathbf{q}+\mathbf{G})} M_{e}(\mathbf{k}, \mathbf{q}+\mathbf{G})\\
&\times\left(d_{\mathbf{k}+\mathbf{q}, e, s}^{\dagger} d_{\mathbf{k}, e, s}-\frac{1}{2} \delta_{\mathbf{q}, \mathbf{0}}\right),
\end{aligned}\label{equ:ooperator}
\end{equation}
where $M_{e}(\mathbf{k}, \mathbf{q}+\mathbf{G})= \alpha_{0}(\mathbf{k}, \mathbf{q}+\mathbf{G}) + ie \alpha_{2}(\mathbf{k}, \mathbf{q}+\mathbf{G})$. 
Following the Lagrange multiplier method introduced in Refs. \cite{PhysRevB.103.205413,PhysRevB.103.205414}, 
the ground state satisfies the equation
$
\left(O_{\mathbf{q}, \mathbf{G}}-A_{\mathbf{G}} N \delta_{\mathbf{q}, 0}\right)|\Psi\rangle=0,
$
where $N$ is the total number of electrons and $A_{\mathbf{G}}$ is the multiplier (see details in Appendix D). Assuming an integer filling $\nu$ and $A_{\mathbf{G}}=\frac{\nu}{N} \sqrt{V(\mathbf{G})} \sum_{\mathbf{k}} \alpha_{0}(\mathbf{k}, \mathbf{G})$, the ground states take the form 
$\left|\Psi_{\nu}^{\nu_{+}, \nu_{-}}\right\rangle=\prod_{\mathbf{k}} \prod_{j_{1}=1}^{\nu_{+}} d_{\mathbf{k},+1, s_{j_{1}}}^{\dagger} \prod_{j_{2}=1}^{\nu_{-}} d_{\mathbf{k},-1, s_{j_{2}}}^{\dagger}|0\rangle,$
where $\nu+2=\nu_{+} +\nu_{-}$ is the total filling factor of the system with $\nu_{\pm}$ being the integer filling of the Chern bands with Chern number $C=\pm2$. It is clear that the state $\left|\Psi_{\nu}^{\nu_{+}, \nu_{-}}\right\rangle$ carries a Chern number of $2(\nu_{+}-\nu_{-})$ and different states with the same $\nu$ are degenerate. 

In real TBCB systems, it is difficult to tune the intrasublattice hopping $w_{AA}$ to be strictly zero. When $w_{AA}\ne0$, neither the particle-hole symmetry nor the chiral symmetry holds (see Appendix C for details). Thus the $M$ matrix takes a general form
\begin{eqnarray}
M(\mathbf{k}, \mathbf{q}+\mathbf{G})&=&\zeta^{0} \alpha_{0}(\mathbf{k}, \mathbf{q}+\mathbf{G}) + \zeta^{x}\alpha_{1}(\mathbf{k}, \mathbf{q}+\mathbf{G})\nonumber\\
&+&i \zeta^{y} \alpha_{2}(\mathbf{k}, \mathbf{q}+\mathbf{G})+\zeta^{z}\alpha_{3}(\mathbf{k}, \mathbf{q}+\mathbf{G}).~~~~
\end{eqnarray}
In the Chern basis, the operator $O_{\mathbf{q}, \mathbf{G}}=O_{\mathbf{q}, \mathbf{G}}^0+O_{\mathbf{q}, \mathbf{G}}^1$ where $O_{\mathbf{q}, \mathbf{G}}^0$ is given by Eq.~(\ref{equ:ooperator}), while $O_{\mathbf{q}, \mathbf{G}}^1$ reads
\begin{eqnarray}
O_{\mathbf{q}, \mathbf{G}}^{1}\!=\!\sum_{\mathbf{k} s,e=\pm 1}  \sqrt{V(\mathbf{q}+\mathbf{G})}
 F_{e}(\mathbf{k}, \mathbf{q}+\mathbf{G}) d_{\mathbf{k}+\mathbf{q},-e, s}^{\dagger} d_{\mathbf{k}, e, s},~~~~~
\end{eqnarray}
where $F_{e}(\mathbf{k},\mathbf{q}+\mathbf{G})\!=\! ie\alpha_{1}(\mathbf{k},\mathbf{q}\!+\!\mathbf{G}) \!+\!  \alpha_{3}(\mathbf{k},\mathbf{q}\!+\!\mathbf{G})$. When the system has an even filling factor $\nu$ and assuming the flat band approximation, the ground state of the operator $O_{\mathbf{q}, \mathbf{G}}^{1}$ becomes $\left|\Psi_{\nu}\right\rangle=\prod_{\mathbf{k}} \prod_{j=1}^{(\nu+2) / 2} d_{\mathbf{k},+1, s_{j}}^{\dagger} d_{\mathbf{k},-1, s_{j}}^{\dagger}|0\rangle$, which has a zero Chern number. For odd filling factors $\nu$, after taking $O_{\mathbf{q}, \mathbf{G}}^{1}$ as a perturbation, the degeneracy of the different Chern states will be lifted, and $O_{\mathbf{q}, \mathbf{G}}^{1}$ prefers the ground state with minimum Chern number. For instance, when $\nu=\pm1$, the system possesses a ground state with broken time-reversal symmetry and high Chern number $C=\pm2$.

{\bf Discussions and concluding remarks:} 
In this paper, we proposed a twisted bilayer system of fermions with $C_4$-symmetry-protected quadratic band touching, which can exhibit exactly flat bands with high Chern numbers $C=\pm2$. Our system's symmetry and the \textit{stable} QBTPs are noteworthy aspects of this study of the twisted graphene system.  The origin of the exactly flat band is related to the anti-holomorphic property of the Hamiltonian in the chiral limit with $t \neq 2 t^{\prime}$. At the first magic angle, the flatness of the topological bands is rather robust against deviation from the exactly flat conditions; that is, the topological bands in the middle exhibiting a high Chern number of $\pm 2$ are nearly flat for a wide range of parameters. See details in the Appendix A.

Such a TBCB lattice may be realized by loading cold atoms into a specially-designed optical-lattice system \cite{Wirth2011,Sun2012}. It has been proposed that the twisted square lattice can be realized by introducing four states (labeled by spin $\pm1/2$ and two layers) and constraining each "layer" by a set of square optical lattices that differ by polarization and a small twisting angle \cite{PhysRevA.100.053604}. If $2\pi$ fluxes are added to the square plaquettes such that the hopping amplitude along its diagonal is $-t'$, the TBCB lattice maybe experimentally realized, and the quantum anomalous Hall effect associated with the flat band with high Chern number may be observed. 
Furthermore, away from integer filling, it is also possible to realize interesting phases such as unconventional superconductivity and fractional Chern insulators, which are left for future studies. 

{\bf Acknowledgments:} 
We would like to thank Andrei Bernevig, Eduardo Fradkin, Steve Kivelson, Kai Sun, and Zhi-Da Song for their helpful discussions. This work was supported in part by the MOSTC under Grant No. 2018YFA0305604 (H.Y.), the NSFC under Grants No. 11825404 (M.-R.L., A.-L.H., and H.Y.) and No. 12204404 (A.-L.H.), and the CAS Strategic Priority Research Program under Grant No. XDB28000000 (H.Y.).




\appendix
  \setcounter{equation}{0}
  \setcounter{figure}{0}
  \setcounter{table}{0}
  \makeatletter
  \renewcommand{\theequation}{S\arabic{equation}}
  \renewcommand{\thefigure}{S\arabic{figure}}

\subsection{Appendix A. Robustness of the Flat Bands}\label{app:robustness}

\begin{figure}[h]
\centering
  \includegraphics[width=1.0\linewidth]{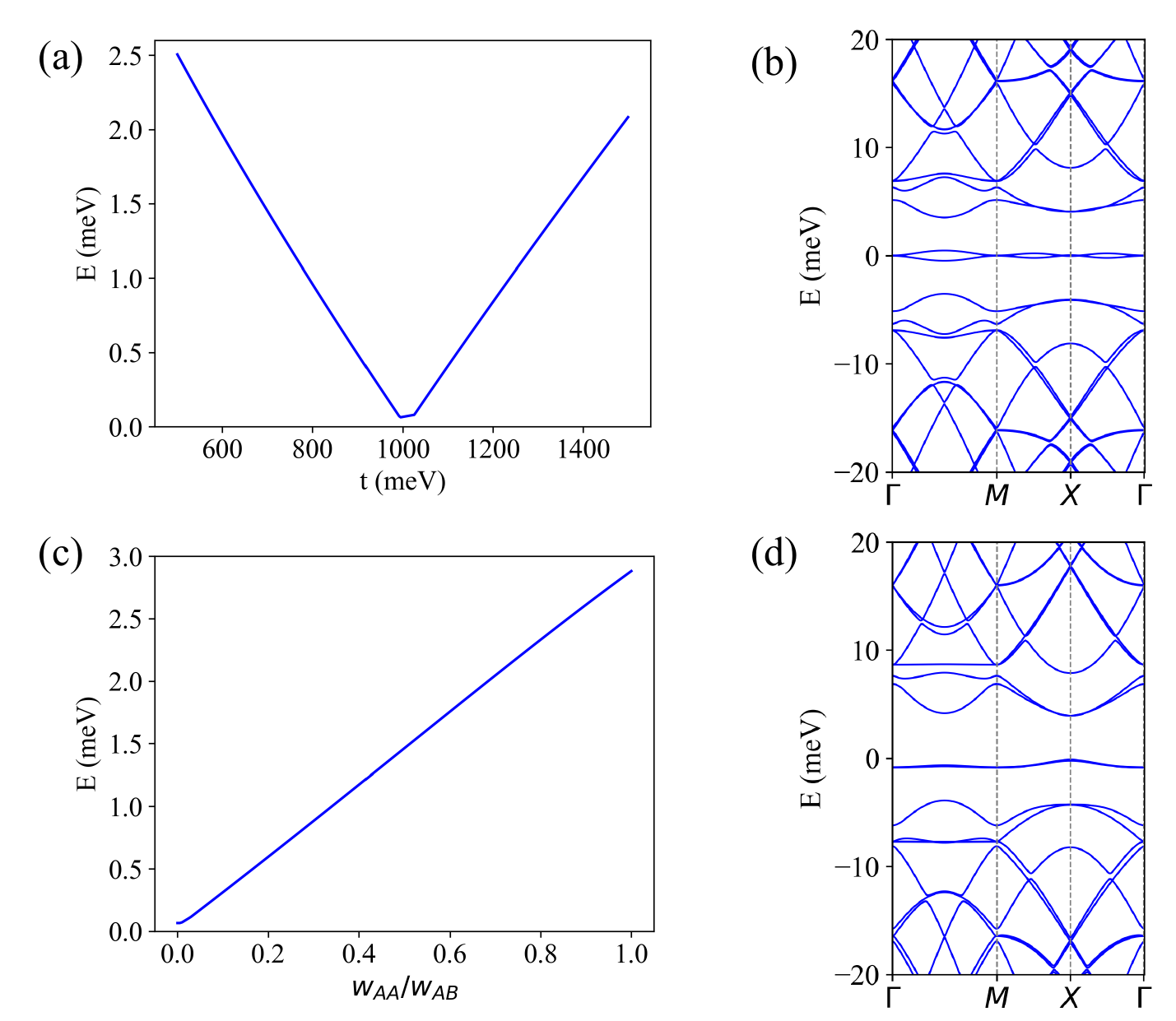}
  \caption{(a) The bandwidth of the middle two flat bands while varying $t$ from 500 to 1500 meV. (b) The band structure with $t^{\prime}=500 \text{ meV}, t=800 \text{ meV}, w_{AA}=0.0 \text{ meV}, w_{AB}=2.05 \text{ meV}, \text{and } \theta=1.6^{\circ}$. (c) The bandwidth of the middle two flat bands while varying $w_{AA}$ from 0 to 2.0 meV. (d) The band structure with $t^{\prime}=500 \text{ meV}, t=1000 \text{ meV}, w_{AA}=0.5 \text{ meV}, w_{AB}=2.05 \text{ meV}, \text{and } \theta=1.6^{\circ}$.}
  \label{fig:robustness}
\end{figure}

As discussed in the main text, the exactly flat band criteria of the TBCB require $t=2t^{\prime}$ and $w_{AA}=0$. In this appendix, we show how the flatness of the two flat bands is affected by these two parameters near the first magic angle. We calculated the bandwidth of the middle two flat bands while varying $t$ or $w_{AA}$. The results are shown in Fig.~\ref{fig:robustness}. Notice that if $w_{AA}\ne0$, the particle-hole symmetry is broken, and thus the flat bands deviate from zero energy.

As shown in Fig.~\ref{fig:robustness}(a), as $t$ varies from $500$ meV to $1500$ meV, the bandwidth of the flat bands varies from 0 to 2.5 meV which is relatively small. As $w_{AA}$ varies from $0$ to $2.0$ meV [Fig.~\ref{fig:robustness}(c)], the bandwidth of the flat bands varies from 0 to 3 meV. The flat bands are quite robust against the deviation of the parameters.

\subsection{Appendix B. Another Derivation of the First Magic Angle}\label{app:derivemagic} 
We now provide another derivation of the first magic angle by requiring the vanishing of the inverse effective mass of the fermions at the QBTP. (Note that this is in contrast to the magic-angle definition of vanishing Fermi velocity in TBG.) Here we only consider the nearest four $M$ points of the bottom layer (red) to the center point of the top layer (blue) in the mBZ, as shown in Fig.~\ref{fig:morielattice} of the main text. We write the ten-band Hamiltonian in the momentum space for these five points in the mBZ:
\begin{equation}
\mathcal{H}(\mathbf{k})=\left(
\begin{array}{ccccc}
\mathcal{H}_0(\mathbf{0}) & \mathcal{T}_1 & \mathcal{T}_2 & \mathcal{T}_3 & \mathcal{T}_4 \\
\mathcal{T}_1 & \mathcal{H}_0(\mathbf{q_1}) & 0 & 0 & 0  \\
\mathcal{T}_2 & 0 & \mathcal{H}_0(\mathbf{q_2}) & 0 & 0  \\
\mathcal{T}_3 & 0 & 0 & \mathcal{H}_0(\mathbf{q_3}) & 0  \\
\mathcal{T}_4 & 0 & 0 & 0 & \mathcal{H}_0(\mathbf{q_4})  \\
\end{array}
\right), 
\end{equation}

where $\mathcal{H}_0(\mathbf{q})\!=\!\left(
\begin{array}{cc}
0 & -it'(\overline{\mathbf{k}}-\overline{\mathbf{q}})^2 \\
 -it'(\mathbf{k}-\mathbf{q})^2 &0  \\
\end{array}
\right)$, $\mathcal{T}_1=\mathcal{T}_3=T_1$, and $\mathcal{T}_2=\mathcal{T}_4=T_2$. The wavefunction $(\psi_t, \psi_{b,1}, \psi_{b,2}, \psi_{b,3}, \psi_{b,4})$ satisfies the Schrödinger equation:
\begin{subequations}
\begin{alignat}{2}
\mathcal{H}_0(\mathbf{0}) \psi_{t}+\sum_{i} \mathcal{T}_{i} \psi_{b, i} &= E \psi_{t}, \\
\mathcal{T}_{i} \psi_{t}+\mathcal{H}_0(\mathbf{q_i}) \psi_{b, i}&=E \psi_{b, i}, \quad i=1,2,3,4.
\end{alignat}
\end{subequations}

From Eq. (S2b) we obtain that $\psi_{b, i}=(E-\mathcal{H}_0(\mathbf{q_i}))^{-1}\mathcal{T}_i\psi_t$, from which we can get the effective Schrödinger equation for $\psi_{t}$:
\begin{equation}
\left[\mathcal{H}_0(\mathbf{0})+\sum_{i}\frac{\mathcal{T}_i\left(E+\mathcal{H}_0(\mathbf{q_i})\right)\mathcal{T}_i}{E^2-t'^2\left(k^2+q_i^2-2\vec{k}. \vec{q}_i\right)^2}\right]\psi_t=E\psi_t.
\end{equation}

Neglecting the $E^2$ and $k^n$ for $n>2$ terms  as small and noticing that $q_i^2=1$, we get
\begin{equation}
\frac{\mathcal{H}_0(\mathbf{0})-\sum_{i}\frac{\mathcal{T}_i \mathcal{H}_0(\mathbf{q_i})\mathcal{T}_i}{t'^2\left(1+k^2-2\vec{k}. \vec{q}_i\right)^2}}{1+\sum_{i}\frac{w^2}{t'^2\left(1+k^2-2\vec{k}. \vec{q}_i\right)^2}}\psi_t=E\psi_t. \label{equ:effha}
\end{equation}

Substituting the $T_{i}$ we have obtained before, one can get the effective Hamiltonian
\begin{equation}
    \mathcal{H}\!=\!\left(
\begin{array}{cc}
0 & -it_{\text{eff}}\overline{\mathbf{k}}^2 \\
 -it_{\text{eff}}\mathbf{k}^2 &0  \\
\end{array}
\right),
\end{equation}
with $t_{\text{eff}}=\frac{1-12\alpha^2}{1+4\alpha^2}t'$. When $\alpha=\frac{1}{\sqrt{12}}\approx0.289$, $t_{\text{eff}}\propto m_{\text{eff}}^{-1}$ tends to zero and flat bands emerge. This result is close to the first magic angle we obtained numerically.\\

\subsection{Appendix C. Symmetries of the Twisted Bilayer Checkerboard Lattice and Gauge fixing}\label{sup:symmetry}

The Hamiltonian of the twisted bilayer checkerboard lattice is
\begin{equation}
\begin{aligned}
H&=\sum_{l}\sum_{\mathbf{k}}f_{l,\mathbf{k}}^{\dagger}h_{l\theta/2}(\mathbf{k})f_{l,\mathbf{k}} \\
&+ \sum_{\mathbf{k}}\sum^2_{i=1}\left(f_{1,\mathbf{k}}^{\dagger}T_{i}f_{-1,\mathbf{k+q_i}}+f_{1,\mathbf{k}}^{\dagger}T_{i}f_{-1,\mathbf{k-q_i}}+\text{H.c.}\right),\label{equ:totalHamCK}
\end{aligned}
\end{equation}
where $h_{l\theta/2}(\mathbf{k})$ is the kinetic term of the checkerboard lattice with a twist angle $l\theta/2$ from the $x$ axis [$l=+1$ $(l=-1)$ for the upper (lower) layer] and has the form of Eq.(1) in the main text. Let $\sigma$ and $\tau$ represent the Pauli matrix for the sublattice degrees of freedom and the layer degrees of freedom respectively. The Hamiltonian with the moiré BZ as shown in Fig.~\ref{fig:morielattice}(b) respects the following spatial symmetries.

$C_{2x/y}$ symmetry
\begin{equation}
        C_{2x/y} f_{\boldsymbol{k}} C_{2x/y}^{-1}= \sigma_{z} f_{C_{2x/y} \boldsymbol{k}}, \quad \left[H, C_{2x/y}\right]=0,
\end{equation}

$C_{4z}$ symmetry 
\begin{equation}
       C_{4z} f_{\boldsymbol{k}} C_{4z}^{-1}=\sigma_{y} f_{R_{\pi/2} \boldsymbol{k}}, \quad \left[H, C_{4z}\right]=0,
\end{equation}


The Hamiltonian also processes the particle-hole symmetry and time-reversal symmetry

Particle-hole symmetry
\begin{equation}
       P f_{\boldsymbol{k}} P^{-1}=\sigma_{y} f_{-\boldsymbol{k}}^{\dagger}, \quad PH(\mathbf{k})P^{-1}=-H^{*}(-\mathbf{k}),\label{equ:particle-hole}
\end{equation}

Time reversal symmetry
\begin{equation}
        T f_{\boldsymbol{k}} T^{-1}=f_{-\boldsymbol{k}}, \quad \left[H, T\right]=0.
\end{equation}
Notice that here the particle-hole symmetry is a rigorous one but will be broken when $w_{AA}\ne0$ which is different from TBG. The system also preserves chiral symmetry if $w_{AA}=0$ with the operator: $\sigma_{y}T$. 

With these symmetries, we can fix the gauge of the wave function. We introduce the sewing matrix $B^{g}(\mathbf{k})$ for the $C_{2z}$, $T$ and $P$ symmetries.
\begin{equation}
\left[D\left(C_{2 z}\right)\right] u_{n}(\mathbf{k})=\sum\left[B^{C_{2 z}}(\mathbf{k})\right]_{m, n} u_{m}(-\mathbf{k}),
\end{equation}
\begin{equation}
\left[D\left(T\right)\right] u_{n}(\mathbf{k})=\sum\left[B^{T}(\mathbf{k})\right]_{m, n} u_{m}(-\mathbf{k}),
\end{equation}
\begin{equation}
\left[D\left(P\right)\right] u_{n}(\mathbf{k})=\sum\left[B^{P}(\mathbf{k})\right]_{m, n} u_{m}(-\mathbf{k}).
\end{equation}

The sewing matrix can be simplified as
\begin{equation}
\begin{aligned}
&{\left[B^{C_{2 z}}(\mathbf{k})\right]_{m, n }= \delta_{m, n} e^{i \varphi_{n}^{C_{2 z}(\mathbf{k})}}}, \\ &{\left[B^{T}(\mathbf{k})\right]_{m, n}= \delta_{m, n} e^{i \varphi_{n}^{T}(\mathbf{k})}}, \\
&{\left[B^{P}(\mathbf{k})\right]_{m, n}= \delta_{-m, n} e^{i \varphi_{n}^{P}(\mathbf{k})} .}
\end{aligned}
\end{equation}

These three symmetry operators can be combined to obtain two independent symmetry operations $C_{2z}P$ and $C_{2z}T$ which keep $\mathbf{k}$ unchanged. The corresponding sewing matrices are defined by the following equations
\begin{equation}
\begin{aligned}
&{\left[D\left(C_{2 z}\right) D(T)\right] u_{n}^{*}(\mathbf{k})=\sum_{m}\left[B^{C_{2 z} T}(\mathbf{k})\right]_{m, n} u_{m}(\mathbf{k}),} \\
&{\left[D(P) D\left(C_{2 z}\right)\right] u_{n}(\mathbf{k})=\sum_{m}\left[B^{C_{2 z} P}(\mathbf{k})\right]_{m, n} u_{m}(\mathbf{k}) .}
\end{aligned}
\end{equation}
The symmetry operations $C_{2z}P$ and $C_{2z}T$ satisfy the properties 
\begin{equation}
\left(C_{2 z} T\right)^{2}=\left(C_{2 z} P\right)^{2}=1, \quad\left[C_{2 z} T, C_{2 z} P\right]=1.
\end{equation}
Thus we can adopt the following $\mathbf{k}$-independent sewing matrices
\begin{equation}
\left[B^{C_{2 z} T}(\mathbf{k})\right]_{m, n}=\delta_{m, n}, \quad\left[B^{C_{2 z} P}(\mathbf{k})\right]_{m, n}=-\operatorname{sgn}(n) \delta_{-m, n}.
\end{equation}
These sewing matrices can also be expressed by the Pauli matrix for the two flat bands
\begin{equation}
B^{C_{2 z} T}(\mathbf{k})=\zeta^{0}, \quad B^{C_{2 z} P}(\mathbf{k})=i\zeta^{y},\label{equ:sewing}
\end{equation}
where $\zeta$ represents the Pauli matrix for the two-flat-band subspace. We have chosen a similar form to the sewing matrix of the TBG systems adopted in Ref.\cite{PhysRevB.103.205413}, and the difference is that TBCB systems do not have two valleys. The wave function and thus the $M$ matrix introduced in the main text and Appendix D,
\begin{eqnarray}
M_{mn}(\mathbf{k}, \mathbf{q}\!+\!\mathbf{G})\!=\!\sum_{\alpha,\mathbf{Q} \in \mathcal{Q}_{\pm}} u_{\mathbf{Q}-\mathbf{G}, \alpha, m}^{*}(\mathbf{k}+\mathbf{q}) u_{\mathbf{Q}, \alpha, n}(\mathbf{k}),~~~~~~~
\end{eqnarray}
are also constrained by the two symmetries, $C_{2z}T$ and $C_{2z}P$, with the sewing matrices we obtained in Eq.~(\ref{equ:sewing}),
\begin{equation}
\begin{aligned}
M_{m n}(\mathbf{k}, \mathbf{q}+\mathbf{G}) 
&=M_{m n}^*(\mathbf{k}, \mathbf{q}+\mathbf{G}) 
,\\
M_{m n}(\mathbf{k}, \mathbf{q}+\mathbf{G}) 
&=\left[\zeta^{y} M(\mathbf{k}, \mathbf{q}+\mathbf{G}) \zeta^{y}\right]_{m, n}.\label{equ:CLconstraint}
\end{aligned}
\end{equation}
Thus the $M$ matrix takes the form
\begin{equation}
M(\mathbf{k}, \mathbf{q}+\mathbf{G})=\zeta^{0} \alpha_{0}(\mathbf{k}, \mathbf{q}+\mathbf{G}) + i \zeta^{y} \alpha_{2}(\mathbf{k}, \mathbf{q}+\mathbf{G}),\label{equ:MmartixinCL}
\end{equation}
where $\alpha_{0}(\mathbf{k}, \mathbf{q}+\mathbf{G})$ and $\alpha_{2}(\mathbf{k}, \mathbf{q}+\mathbf{G})$ are real numbers. Besides, from the definition of the $M$ matrix in the main text, the $M$ matrix also satisfies the Hermiticity condition
\begin{equation}
    M_{m n}^*(\mathbf{k}, \mathbf{q}+\mathbf{G}) 
=M_{n m}(\mathbf{k}+\mathbf{q}, -\mathbf{q}-\mathbf{G}),
\end{equation}
which means that $\alpha_{i}(\mathbf{k}, \mathbf{q}+\mathbf{G})$ satisfy
\begin{equation}
\begin{aligned}
\alpha_{0}(\mathbf{k}, \mathbf{q}+\mathbf{G})&=\alpha_{0}(\mathbf{k}+\mathbf{q},-\mathbf{q}-\mathbf{G}) \\
\alpha_{2}(\mathbf{k}, \mathbf{q}+\mathbf{G})&=-\alpha_{2}(\mathbf{k}+\mathbf{q},-\mathbf{q}-\mathbf{G}).\label{equ:clalphaconstraint0}
\end{aligned}
\end{equation}

We can also fix the relative gauge between the wave functions with momentum $\mathbf{k}$ and those with momentum $-\mathbf{k}$ by $C_{2z}$ symmetry. Notice that in the TBCB system, $C_{2z}$, $T$, and $P$ symmetries commute with each other, and we can choose the sewing matrix for these three symmetries
\begin{equation}
B^{C_{2 z}}(\mathbf{k})=\zeta^{0}, \quad B^{T}(\mathbf{k})=\zeta^{0}, \quad B^{P}(\mathbf{k})=i\zeta^{y}.\label{equ:sewing2}
\end{equation}
Thus, the $M$ matrix also has the following constraint implied between the momentum $\mathbf{k}$ and the momentum $-\mathbf{k}$:
\begin{equation}
M_{m n}(\mathbf{k}, \mathbf{q}+\mathbf{G}) 
=M_{m n}(-\mathbf{k}, -\mathbf{q}-\mathbf{G}),\label{equ:CLconstraint1}
\end{equation}
which implies that
\begin{equation}
\alpha_{a}(\mathbf{k}, \mathbf{q}+\mathbf{G})=\alpha_{a}(-\mathbf{k},-\mathbf{q}-\mathbf{G})\quad \text{ for } a =0,2.\label{equ:clalphaconstraint1}
\end{equation}

When the hopping $w_{AA} \ne 0$, the particle and chiral symmetries are broken, and Eq.~(\ref{equ:CLconstraint}) no longer holds. Constrained by the real condition, the $M$ matrix takes a more general form,
\begin{equation}
\begin{aligned}
M(\mathbf{k}, \mathbf{q}+\mathbf{G})&=\zeta^{0} \alpha_{0}(\mathbf{k}, \mathbf{q}+\mathbf{G}) + \zeta^{x}\alpha_{1}(\mathbf{k}, \mathbf{q}+\mathbf{G})\\
&+i \zeta^{y} \alpha_{2}(\mathbf{k}, \mathbf{q}+\mathbf{G})+\zeta^{z}\alpha_{3}(\mathbf{k}, \mathbf{q}+\mathbf{G})\\
&=M_{0}(\mathbf{k}, \mathbf{q}+\mathbf{G})+M_{1}(\mathbf{k}, \mathbf{q}+\mathbf{G}),
\end{aligned}
\end{equation}
where $\alpha_{i}(\mathbf{k}, \mathbf{q}+\mathbf{G})$ ($i=0,1,2,3$) are real numbers.

Similar to the chiral case introduced above, now $\alpha_{i}(\mathbf{k}, \mathbf{q}+\mathbf{G})$ are also constrained by the Hermiticity condition and the $C_{2z}$ symmetry:
\begin{equation}
\begin{aligned}
\alpha_{a}(\mathbf{k}, \mathbf{q}+\mathbf{G})&\alpha_{a}(\mathbf{k}+\mathbf{q},-\mathbf{q}-\mathbf{G}) \quad \text { for } a=0,1,3,\\
\alpha_{2}(\mathbf{k}, \mathbf{q}+\mathbf{G})&=-\alpha_{2}(\mathbf{k}+\mathbf{q},-\mathbf{q}-\mathbf{G}),
\end{aligned}
\end{equation}
\begin{equation}
\alpha_{a}(\mathbf{k}, \mathbf{q}+\mathbf{G})=\alpha_{a}(-\mathbf{k},-\mathbf{q}-\mathbf{G}),\quad \text{ for } a =0,1,2,3.
\end{equation}

\subsection{Appendix D. Solving the Ground State of the Interacting Hamiltonian}\label{sup:interactingH}

The Coulomb interacting Hamiltonian of the system in the momentum space is written as
\begin{equation}
\mathcal{H}=\frac{1}{2 A} \sum_{\mathbf{G} \in \mathcal{G},\mathbf{q} \in \mathrm{mBZ}} V(\mathbf{q}+\mathbf{G}) \delta \rho_{-\mathbf{q}-\mathbf{G}} \delta \rho_{\mathbf{q}+\mathbf{G}},\label{equ:clham}
\end{equation}

where the gate Coulomb potential is $V(\mathbf{q})=\pi d^{2} U_{d} \frac{\tanh (d|\mathbf{q}| / 2)}{d|\mathbf{q}| / 2}$. Under the Chern band basis, the charge density term $\delta \rho_{-\mathbf{q}-\mathbf{G}}$ is 
\begin{equation}
\delta \rho_{\mathbf{G}+\mathbf{q}}=\sum_{\mathbf{k}, s} \sum_{m, n} M_{m, n}( \mathbf{k},\mathbf{q}+\mathbf{G})\bigg(c_{m, s}^{\dagger}(\mathbf{k}+\mathbf{q}) c_{n, s}(\mathbf{k})
-\frac{1}{2} \delta_{\mathbf{q}, 0} \delta_{m n}\bigg),
\end{equation}
where 
\begin{equation}
M_{m, n}(\mathbf{k}, \mathbf{q}+\mathbf{G})=\sum_{\alpha} \sum_{\mathbf{Q} \in \mathcal{Q}_{\pm}} u_{\mathbf{Q}-\mathbf{G}, \alpha, m}^{*}(\mathbf{k}+\mathbf{q}) u_{\mathbf{Q}, \alpha, n}(\mathbf{k}).
\end{equation}

The interacting Hamiltonian now is written in a semi-positive definite form
\begin{equation}
\mathcal{H}=\frac{1}{2 \Omega_{\text {tot }}} \sum_{\mathbf{q} \in \text { mBZ }} \sum_{\mathbf{G} \in \mathcal{G}} O_{-\mathbf{q},-\mathbf{G}} O_{\mathbf{q}, \mathbf{G}},
\end{equation}

where $O_{\mathbf{q}, \mathbf{G}}=\sum_{\mathbf{k} s} \sum_{m, n=\pm 1} \sqrt{V(\mathbf{q}+\mathbf{G})} M_{m, n}(\mathbf{k}, \mathbf{q}+\mathbf{G}) \left(c_{m, s}^{\dagger}(\mathbf{k}+\mathbf{q}) c_{n, s}(\mathbf{k})-\frac{1}{2} \delta_{\mathbf{q}, 0} \delta_{m, n}\right)$.

Notice that the number of the electron $N$ is conserved; thus we are able to introduce a Lagrange multiplier $A_{\mathbf{G}}$
\begin{widetext}
\begin{equation}
\mathcal{H}=\frac{1}{2 \Omega_{\mathrm{tot}}} \sum_{\mathrm{G} \in \mathcal{G}}\left[\left(\sum_{\mathrm{q}}\left(O_{\mathrm{q}, \mathrm{G}}-A_{\mathrm{G}} N \delta_{\mathbf{q}, 0}\right)\left(O_{-\mathrm{q},-\mathrm{G}}-A_{-\mathrm{G}} N \delta_{-\mathbf{q}, 0}\right)\right)+2 A_{-\mathrm{G}} N O_{0, \mathrm{G}}-A_{-\mathrm{G}} A_{\mathrm{G}} N^{2}\right].\label{equ:LMHam}
\end{equation}
\end{widetext}
When the flat metric condition \cite{PhysRevB.103.205413,PhysRevB.103.205414} $M_{m, n}(\mathbf{k}, \mathbf{G})=\xi(\mathbf{G}) \delta_{m, n}$ is satisfied or filling factor $\nu=0$, the last two terms in Eq.~(\ref{equ:LMHam}) are constant which depend on $N$. In this way, one can easily conclude that the ground state of the interacting Hamiltonian satisfies the equation
\begin{equation}
\left(O_{\mathbf{q}, \mathbf{G}}-A_{\mathbf{G}} N \delta_{\mathbf{q}, 0}\right)|\Psi\rangle=0.\label{equ:groundequ}
\end{equation}

To solve the ground state, one only needs to solve Eq.~(\ref{equ:groundequ}). In general, the Flat Metric Condition is not strictly satisfied except for $\mathbf{G}=\mathbf{0}$. Fortunately, when the flat metric condition is not largely violated, the ground states which satisfy Eq.~(\ref{equ:groundequ}) persist as long as the gap between the ground states and exciting states is not closed. Since the wave function decreaseS exponentially as $\mathrm{G}$ increases for the moiré Hamiltonian\cite{PhysRevB.103.205411}, one can assume that the flat metric condition is not largely violated and the ground state derived above is the real ground state of the system. Future studies can adopt the real-space projection method \cite{PhysRevX.8.031088,PhysRevLett.122.246401} to confirm our conclusion for ground states.

\bibliography{ArXiv_version}

%
%

\end{document}